# Garment Suggestion Based on Comfort Extracted from Physiological and Emotional Parameters

Hyo Jung (Julie) Chang, *Department of Hospitality and Retail Management, Texas Tech University, Lubbock, Texas, USA,* Mohammad Abu Nasir Rakib, *Department of Computer Science and Engineering, University of Texas at Arlington, Arlington, Texas, USA,* Md Kamrul Hasan Foysal, *Department of Electrical and Computer Engineering, Texas Tech University, Lubbock, Texas, USA*, and Jo Woon Chong, *Department of Electrical and Computer Engineering, Texas Tech University, Lubbock, Texas, USA*

## 1. Introduction

Garment comfort is a complex term that is related to several factors including fit, fiber type, and style, which could also be evaluated differently from a person to person (Zimniewska & Kozłowski, 2004). Different types of fibers may affect the comfortability of the garments in a different way. In general, clothing made from natural fibers, like cotton, provide more comfortability than clothing made from synthetic fibers (Cook, 1984). Especially, natural fibers including cotton and linen are more breathable than synthetic fibers, thus, wearers are more likely to feel comfortable while wearing the fabrics made from these fibers (Aisyah & Tajuddin, 2014). Recent advancements in textile finishing showed scopes of improving comfortability, however, synthetic fibers are still less comfortable because of lower moisture absorbency, heat conductivity, and sticking out with the body in dry seasons, among others (El Nemr, 2012). Along with aesthetic features, thermal comfort and breathability from a textile product would benefit the wearer to maintain a healthy lifestyle. Furthermore, previous research demonstrated that polyester fibers cause higher body temperature, more discomfort, and sweating during summer or high-temperature environments (Kwon, et. al., 1998). Thus, based on the environmental condition, wearers' activities and their emotional conditions, the overall physiological comfort of a person can be different.

Especially, cotton products provide high breathable properties and moisture absorbency properties (Cook, 1984), thus, it is important to determine whether these breathable and moisture absorbency properties have an overall impact on the wearer's physiological responses to measure the comfort, such as heart rate and respiration rate. Ciesielska et. al. (2009) showed both natural and man-made fiber-based garments have an influence on the wearer's physiological states such as cardiovascular and respiration rate during physiological effort. Even though garments produced from natural fibers create more positive feeling of the wearers, wool-based garments are not suitable after the works that require high physical activities (Ciensielska, 2009). Synthetic fibers such as polyester do not provide as much moisture regains as cotton and it generates static electricity, however, it is unknown whether these characteristics of natural and synthetic fiber fabric have any impact on the wearer's emotional state.

Apart from the fabric and fiber types, a garment fit is one of the important factors that have an impact on the physical comfort of a human body. This physical comfort is assumed to be related to the emotional state of the wearers. Nowadays, consumers prefer to wear garments with various fit types, such as loose fit, regular fit, or tight fit. However, no study as of today confirmed the effect of garment fit on the wearer's physiological parameters, which are closely related to their health. Apart from the respiration rate and heart rate, the garment's surface smoothness and roughness could be another potential factor that might influence wearers' comfortability (Ciensielska, 2009). For example, wool is itchy because of having scale on its surface structure, so that the wearer may feel uncomfortable while worn (Hearle & Morton, 2008). As a result, negative emotions may be induced for the wearer. It can be assumed that consumers may experience emotionally connected if they find good hand-feel, accurate and desired to fit, and non-bulky feeling while worn. On contrary, synthetic fibers are more likely to produce static electricity, which is undesired by the wearers. Thus, the research question of this study was "Do the physiological and emotional responses, as the measurements of the comfortability, of the wearers may vary based on the fabric, comfortability, and fit types?" By monitoring heart rate and respiration rate in addition to the effect of fabric types and fit on the wearer's emotion would disclose what fabric and fit type a person should consider in their daily life. Thus, the objectives of this study are (1) to evaluate the effect of garment fitting and fabric types on heart rate and breathing rate; (2) to determine which clothing can provide better heart and respiration rates between garments made with natural and synthetic fabrics; and (3) to examine the emotional responses toward each fabric type and garment fit.

**2. Background of the study**

*2.1 Physiological responses to clothing*

Previous studies showed mixed results about the effects of fabric types on the physiological comfort of the wearer (Hassan, 2012; Dehghan et. al., 2014; Ciensielska, 2009). For example, Hassan et al. (2012) found that polyester fabrics provide better physiological responses i.e., better cardiorespiratory performance and fitness for the athletes. On other hand, Dehghan et. al., (2014) conducted a study on working-class people to see the effect of fabric types on cardiological and physiological responses. The result showed there was not any significant impact of fabric types on the physiological responses of the participants (Parvari et. al., 2014). Additionally, respiration rate provides data about an individual's health condition (Rolfe, 2019). According to Nicolo et. al., (2020), respiration rate provides some crucial information about the human body, such as cardiac situation and clinical deterioration. Furthermore, respiration rate is also an indicator that provides information about cognitive load, emotional stress, and among others (Nicolo et. al., 2020). Researchers widely use the respiration rate measurement technique, which is a non-invasive test that indicates changes in oxygen levels in the blood as well as breathing difficulties (Cooper et. al., 2014). Moreover, heart rate measurement has been studied to check the endurance of the different athletes and their physical conditions (Gajda, 2020). The result emphasized the importance of continuous monitoring of heart rate, which may improve the safety of the athletes.

*2.2. Emotional responses to garments*

Previously Kasambala et. al. (2016) showed inaccurate garment fit affects individuals' emotional responses from which individuals develop their personal values to achieve certain goals (Kasambala et. al., 2016). Likewise, garment fit, and appearance represent an individual's state of choice that communicates personal attitudes and values (Kaiser, 1998; Kim & Damhorst, 2010). Furthermore, Furthermore, people use emotion as a way of communicating and maintaining social interaction. An emotional response is a part of affective social communication (Penalp, 1999). People communicate emotion through the human body including facial expressions, gestures, and body movements, where the garment type and fit play a significant role in effectively communicating the desired emotion (Mura, 2008). Thus, the garment becomes another mode of communication. Although the expressing emotion through garments is not uniform in people and contains meaningful messages to the wearers and viewers in a different culture. As a result, people tend to exhibit their personality, status, and identity by what they wear (Kasambala et. al., 2016). This emotional expression through clothing may be different based on culture, gender, functionality, and age. For example, Kasambala et al. (2016), conducted a study on South African women emphasizing the relation of emotion, personal values, and garment fit. Their results showed how manufacturers are failing in addressing the right fit and sizing of the garments for the females.

*2.2 Garment fabric and comfort*

In general, people tend to touch the garment product to evaluate the touch and quality of the fabric before buying (Rahman, 2012). Thus, based on the feelings or emotional responses generated due to the interaction from skin contact with garment items help determine whether the consumer would buy that garment product. Furthermore, many investigators pointed out that emotional responses from consumers play an important role in deciding what they buy (Lin, Akimobo, & Shigeo, 2008; Clore, Palmar, & Gratch, 2009). Consumers also experience an interaction with the body while wearing a clothing item. Therefore, fits and surface roughness/smoothness of garment items can enhance the sensorial comfort, subsequently, creates an emotional connection to the garment product (De Raeve, et al., 2018). For example, a soft or smooth surface provides a pleasant experience while a rough or stiff surface provides a less pleasant experience (Essick et. al., 1999). Furthermore, gender, skin contact, texture, fabric constructions, presence of spandex, among others can influence the user experience (Essick, et. al., 2010; Rakib et. al., 2020). However, garment items made from natural fibers i.e., cotton, flax, and silk, are more comfortable to the wearers than garments from synthetic fibers i.e., polyester, nylon, among others (Nemr, 2012). This is due to natural fibers provide the features including amicable to the skin, breathability, higher moisture absorbency, and good thermal conductivity, among others. On contrary, synthetic fibers do not possess properties for which natural fibers are more favorable to the wearers (Annapoorani, 2018). The above-mentioned properties also determine the comfortability of the wearers. However, because of technological advancement, synthetic fibers are possible to modify and provide better properties likewise natural fibers (Parvinzadeh, 2012).

*2.3 Garment fit and comfort*

Along with the fabric types (natural or synthetic), the garment fit also determines the comfortability. As mentioned earlier, comfortability is the harmony of psychological, physical, and physiological states of a human body (Raeve et. al., 2018). The comfortability of a human body is related to several factors including skin contact, air and moisture permeability, fit, fabric handle properties, and among others (Hooper et. al., 20150. Aesthetically, if a garment looks beautiful on a person, it indicates the well-fitting of that garment to the wearer (Alexander et. al., 2005). Therefore, before wearing a garment, a person needs to decide which garment (natural or synthetic) would provide more comfort along with positive emotional feelings. However, there are not many studies conducted to examine the relationships among garment fit, emotional responses, and comfortability (Wu, et. al, 2011).

Different consumers prefer different types of clothing fits i.e., tight fit, loose fit, slim fit, and among others. The garment fit affects the consumer's physiological, emotional, and physical state of a human body. Physiological comfort may be defined as the condition of the human body where a human can get sufficient air, thermal, and moisture regulating ability, among others after wearing a garment (De Raeve et. al., 2018). The fit can also affect the insulation property of garment products. Therefore, based on the season, clothing fit and fabric types determine the thermal insulation and comfort of a human body (Wang, et. al., 2016). The physiological comfort is influenced by several factors including style, fabric types, and the fit of the garments

(Mert, et al., 2016). Also, it is probable that fabric types and fits could affect some physiological responses i.e., respiration rate, heart rate, and body temperature of the wearer.

Therefore, as mentioned earlier, it is important for knowing how the fit and fabric type can help the user maintain comfortability through fashion products. Furthermore, no previous studies disclosed the effect of the garment fit on a wearer's emotional response and physiological responses i.e., heart rate and respiration rate. Therefore, this study is aiming to understand how fabric types, i.e., natural and synthetic products, and garment fit types can affect a wearer's physiological responses, so that the wearers of those fashion products will be aware of selecting a garment product to achieve good breathable features to themselves.

The comfort of the garment can be extracted from physiological parameters (Das & Alagirusamy, 2010). However, no existing fit detection method have measured the physiological responses to the garment. Conventional fit detection approaches complete focus on body measurement and suggest clothing based on only one parameter. As these solutions do not consider comfort and physiological indicators, they do not provide accurate garment suggestion to the consumer. Emotional response to those garments is also important indicator for comfort as positive emotions evoke when an individual feels comfortable when wearing a certain garment. Thus, our proposed method incorporates physiological data (i.e., heart rate and respiration rate) with emotional data to present the best garment fit to the consumer. For all these physiological measurements, no external device is needed. Smartphone camera is used to detect heart rate (Kwon, 2012) and respiration rate (Reyes, 2016) of the consumer.

## 3. Method

In aiming to provide a comprehensive understanding of how fit and comfortability are interrelated and show if there is an overlap exists among the garment fitting, comfortability, and emotional state of the wearers through the extraction of physiological data. Therefore, this study collected physiological data using pulse oximeter and respiration sensors (Nexus 10 MK-II) and emotional data using a self-administered questionnaire based on qualitative data. In our proposed method, we focused on two physiological measurements as comfort parameters i.e., heart rate (HR) and respiration rate (RR). Numerous studies have relied on heart rate (Lan et. al., 2020) and respiration rate (Song, 2011) as comfort parameters. Furthermore, the quantitative survey method was used to measure the emotional responses to the fabric type and fit. Several items were used to evaluate the emotional state and comfortability of the wearers. Additionally, several open-ended questions were also included in the survey.

**3.1 Physiological Parameter Research Design**

Previously, research have shown strong correlation between physiological parameter and clothing comfort. Especially, heart rate, respiration rate and skin conductance are a few parameters that have been assessed to be closely correlated to comfort. In our study, we have taken physiological data (heart rate and respiration rate) as comfort parameters to measure clothing comfort. A gold standard Nexus MK-10 device has been used for acquiring ECG and respiration signal. The ECG and respiration signal were used to extract the heart rate and respiration rate of the participants. Different noise reduction techniques were used to reduce signal noise. The collected data were analyzed and processed using MATLAB.

In this study, we have focused on the following two aspects, namely,

1. Finding direct relationship between comfort parameter and fabric type and fit.
2. Providing garment suggestion based on acquired comfort parameter.

*3.1.1. ECG signal acquisition*

Electrocardiogram (ECG) reflects the electrical activity of heart. A typical ECG signal contains p wave, QRS complex and T wave. Heart rate can easily be acquired from an ECG signal. An ECG machine contains electrodes which measures external (skin) electrical activity of heart. The four extremity electrodes are left arm, right arm, neutral, on the right leg, and foot, on the left leg. However, two electrodes on two arms and a virtual neural can be used to get one lead ECG accurately. There is no difference in acquired data if the electrodes are attached proximal or distal on the extremities. Figure 1 below shows (a) an ECG signal, and (b) ECG signal acquisition process using NEXUS -10 MK-II device.

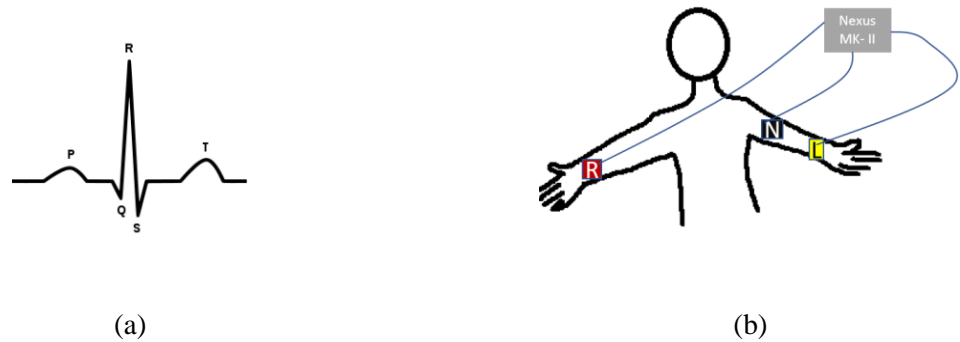

(a) (b)

Figure 1. (a) A typical ECG signal, and (b) ECG signal acquisition using Nexus -10 MK-II device.

For finding the relationship of comfort parameter to fabric type and garment fit, we used gold standard medical device Nexus- 10 MK -II to obtain heart rate and respiration rate from the subjects. Standard electrode was used to obtain the signals from the subjects. Figure 2 shows the Nexus- 10 MK -II device and data acquisition of one subject.

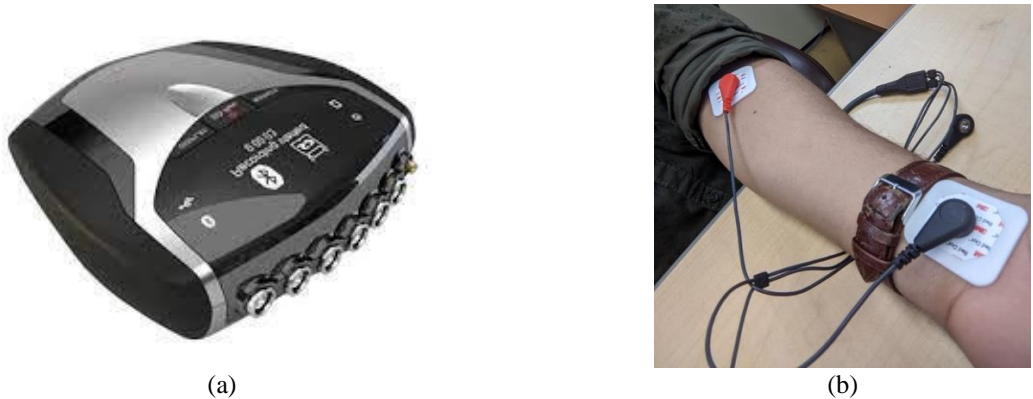

(a) (b)

Figure 2. (a) Nexus 10 MK-II device, and (b) Heart rate and Respiration rate acquisition

*3.1.2 Heart rate calculation*

From the acquired biomedical signal (ECG), the heart rate is extracted. First, noise reduction techniques are applied, and the signal is preprocessed. Low pass filtering, moving average, detrending are some of the techniques that are applied to process the signal. A peak finding algorithm is used to extract the R peaks of the signal. The block diagram in figure 3 shows the heart rate acquisition workflow.

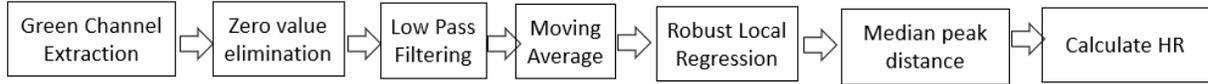

Figure 3. ECG signal processing technique.

The signal processing and extracted signal peaks were demonstrated in the Figure 4.

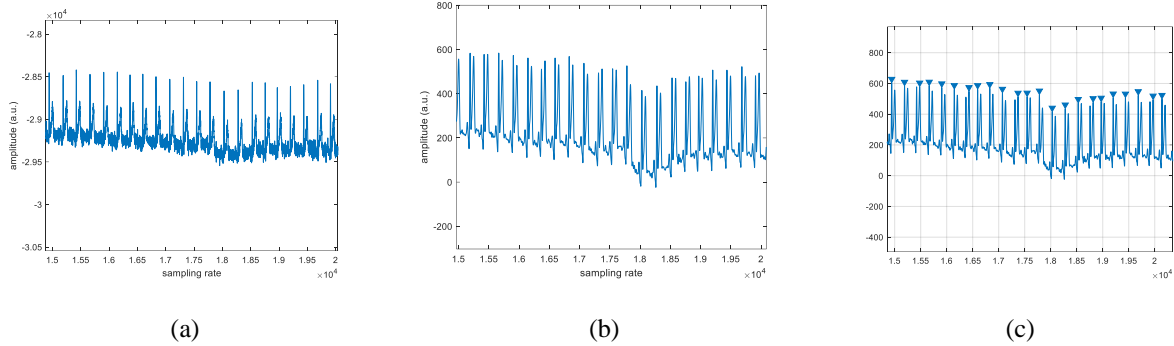

(a)            (b)            (c)

Figure 4. (a) raw ECG signal, (b) processed denoised signal, and (c) final peak detected signal.

Heart rate is calculated using the following equation.

$$R = \frac{60}{t_{p-p}} \quad (1)$$

where $HR$ is the heart rate and $t_{p-p}$ is the peak to peak time difference

### 3.1.3. Respiration signal acquisition

For extraction of respiration signal, the same Nexus 10 MK II device is used. The respiration sensor consists of a piezoelectric transducer which uses expansion and contraction of a piezoelectric band to obtain the respiration signal of the user. Figure 5 below shows the respiration signal extraction method.

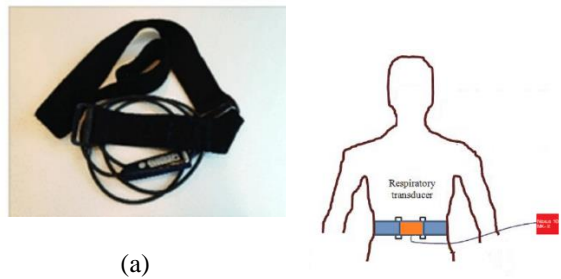

(a)

Figure 5. (a) Respiration Transducer and (b) Experimental Set up to obtain respiration signal.

From the acquired biomedical signal (Respiration Singla), the breathing rate is extracted. First, noise reduction techniques are applied, and the signal is preprocessed. Low pass filtering, moving average, detrending are some of the techniques that are applied to process the signal. A peak finding algorithm is used to extract the peaks of the signal. The block diagram in figure 6 shows the signal processing.

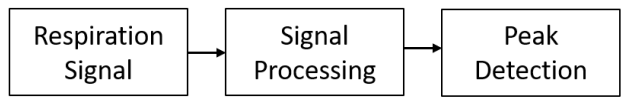

Figure 6. Respiration signal processing and respiration rate extraction technique.

The signal processing and extracted signal peaks were demonstrated in the Figure 7.

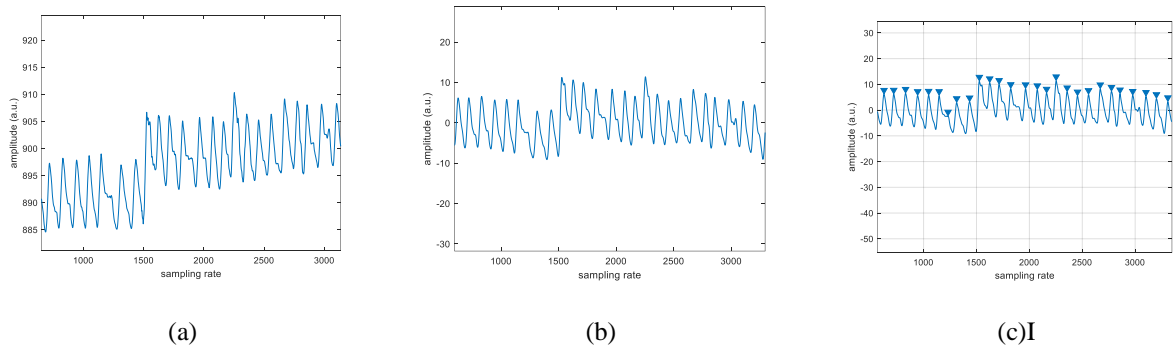

Figure 7. (a) raw signal, (b) processed denoised signal, and (c) final peak detected signal

*3.1.3 Respiration rate calculation*

From the respiration signal, the respiration rate is acquired using equation 2.

$$RR = \frac{60}{t_{p-p}}, \qquad (2)$$

where $RR$ is the respiration rate and $t_{p-p}$ is the peak to peak time difference.

*3.2. Relationship between fabric type and comfort parameter*

Studies have shown direct and indirect relationship of fabric type to human comfort. The parameters used to measure human comfort are heart rate, blood volume pulse, respiration rate and skin temperature. In our study, we relied upon heart rate and respiration rate for comfort parameters.

Clothing fabric type has been the key issue of comfort for the consumers. Empirically, clothing type based on the fabric was considered the key factor for choosing clothing. People have chosen cotton, polyester, linen, and numerous other fabrics to make clothing's. Studies have shown comparison in comfort percipience based on fabric type. Mostly, the main two type of fabrics generally used are cotton and polyester. In this study, we focused on these two fabrics to measure comfort of the subjects. Figure 8 below shows the relation between fabric type and comfort parameter (for visual only).

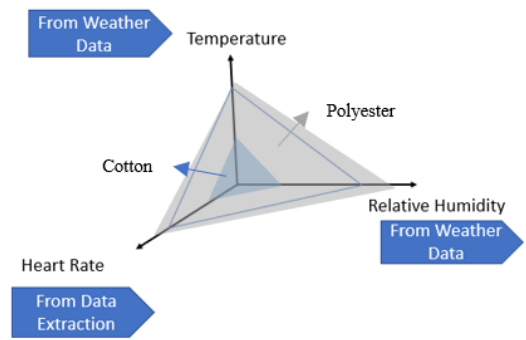

Figure 8. Relationship parameter of comfort and fabric type

*3.3 Relationship between clothing fitting and comfort parameter*

The other issue which shapes our comfort is the fitting condition. Although, the fitting preference depends upon personal preference, there is a baseline for fitting condition. In this study, we focused on fitting criteria i.e., fitted and loose-fitting condition for the clothing to the subjects and extracted the correlation towards them.

*3.4. Finding direct relationship among comfort parameter, fabric type, and garment fit*

The hypothesis we wanted to test was that there is no evidence of Difference in heart rate and respiration rate between two fabric type, two fitting condition, namely, Cotton Tight-Fit, Cotton Loose-Fit, Polyester Tight-Fit, and Polyester Loose-Fit. In order to select the garment for the users, we considered two type of fabric and two types of fitting condition. This study initially thought to use 100% cotton and 100% polyester tee-shirt. However, based on the availability and frequently purchase tee-shirt types, the selected product specification is given in Table 1 below.

Table 1. Parameters for fabric selection

| Fabric Type | Fabric Construction | Fitting Condition |
|---|---|---|
| Cotton- 90% cotton+10 % polyester | Single Jersey Knitted Fabric | Tight-fit |
| Cotton- 90% cotton+10 % polyester | Single Jersey Knitted Fabric | Loose fit |
| Polyester- 90% polyester + 10 % spandex | Single Jersey Knitted Fabric | Tight-fit |
| Polyester- 90% polyester + 10 % spandex | Single Jersey Knitted Fabric | Loose fit |

The fabric type is obtained from the following Table 2.

Table 2. Relationship between Fabric types and comfort parameters according to literature.

| Fabric Type | Heart Rate | Temperature | Humidity | Reference |
|---|---|---|---|---|
| Hydrophilic Cotton | Low | - | - | (Liya et al. 2007) |
| Moisture Management Cotton | High | | | |
| 100% cotton | Low | High | High | (Parvari, Aghaei et al. 2015) |
| 13.7% viscose 86.3% polyester | High | High | High | |
| 30.2% cotton, 69.8% polyester | Low | High | Low | |
| 13.7% viscose 86.3% polyester | High | High | Low | |
| 100% polyester (low moisture regain) | High | 30C | 50% | (Kwon, Kato et al. 1998) |
| Wool + cotton (high moisture regain) | Low | | | |
| 100% cotton (moderate moisture regain) | Low | | | |
| 100% cotton | Low | - | - | (Li, Keighley et al. 1988) |
| 65% polyester, 35% cotton | Low | | | |
| 100% polyester | High | | | |
| Experimental Clothing- Under Armor | Low | - | - | (Wickwire, Bishop et al. 2007) |
| Cotton | High | | | |

*3.5. Providing garment suggestion based on acquired comfort parameter using smartphone app*

In this paper, we propose a novel non-contact garment suggestion method to the online apparel customers. The non-contact smartphone-based method uses video data acquired from customers end before purchase and provides garment suggestion based on physiological comfort parameters. Current smartphone technology is advanced in terms of camera specification. Therefore, it is possible to acquire heart rate and respiration rate of a subject's facial video using smartphone camera. In this specific study, we used smartphone camera to obtain heart rate and respiration rate using smartphone and developed a model to provide optimal garment suggestion to the customers. The proposed model is described in the following section.

*3.5.1 Video acquisition*

Smartphone camera video is used to obtain imaging Photoplethysmogram (ippg) signal (Zhang et. Al., 2008). In this study, we obtained subject's heart rate from face video ippg signal. Smartphone camera obtains video is rgb format, where *R* is for channel Red, *G* for Channel Green and *B* for channel Blue. Therefore, each frame of the video is essentially an image of *RGB* format.

In the proposed algorithm, each frame is extracted from the face video to process and obtain the heart rate.

*3.5.2 Heart Rate and Respiration Rate acquisition*

Facial video has been used earlier to obtain ippg signal in order to extract heart rate and respiration rate accurately (Sanyal & Nundy, 2018; Favilla, et. al., 2018). In our proposed technique, we incorporate this method using smartphone camera for a more convenient user platform for online apparel customers. iPPg is Photoplethysmogram signal obtained from image. The basig of Photoplethysmogram signal acquisition uses a light source and a photodetector. The light source emits light to a tissue and the photodetector measures the reflected light from the tissue. The reflected light is proportional to blood volume variations. Usual PPG sensors use an infrared light emitting diode (IR-LED) or a green LED as the main light source. The green light sensor is used for calculating the absorption of oxygen in oxygenated blood and deoxygenated blood (Peng et. al., 2014; Tabei et. al., 2019). A photodetector is used to measure the intensity of reflected light from the tissue. The blood volume changes can then be measured (calculated) based on the amount of the detected light. iPPG works on the same principle to obtain the blood volume pulse. Figure 9 shows the ppg signal extraction from fingertip.

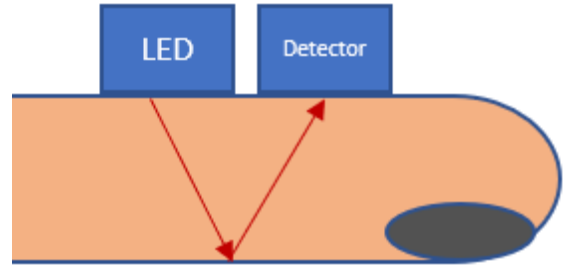

Figure 9. PPG signal extraction.

*3.5.3. Proposed smartphone app for suggesting garment:*

In our method, we developed a smartphone application to automatically obtain the facial video of the customer in front of the smartphone front camera. The video obtained is converted into ippg signal and subsequently heart rate and respiration rate (as shown in (Sanyal & Nundy, 2018)). Based on the heart rate, respiration rate and other parameters (i.e., temperature, location) the user is suggested a fabric type suitable for him/her. The block diagram in Figure 10 below shows the method.

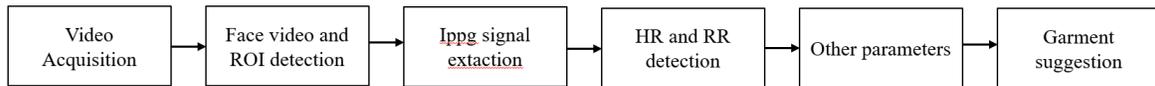

Figure 10. Block diagram for proposed fabric and fit suggestion smartphone app.

Figure 11 below shows the process of obtaining physiological parameters and garment suggestion process.

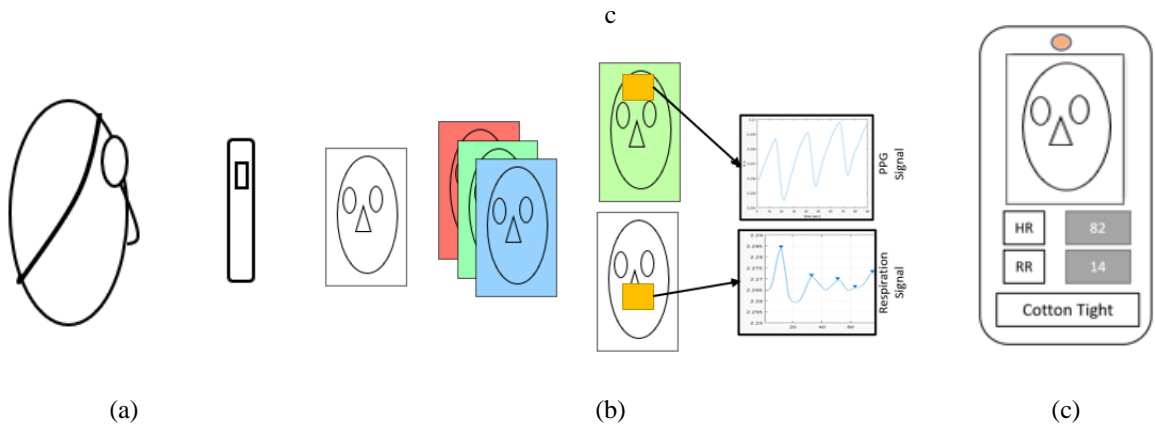

(a)           (b)           (c)

Figure 11. (a) video acquisition, (b) HR and RR extraction (ROIs represented by yellow boxes), and (c) Garment fabric and fit Suggestion.

## 4. Results

*4.1 Data Collection*

For fabric type and fit for comfort parameter, 11 The subjects were recruited including 6 males participants and 5 females following the Texas Tech institutional review board (IRB 2020-482). The clothing condition were: 1) Polyester-Loose Fit, 2) Polyester-Tight Fit, 3) Cotton-Loose Fit, 4) Cotton-Tight Fit. The subjects were asked to use the fitting room (Figure 12 (a)) to change the clothing. After arriving in the experiment setting, the subjects were asked to relax for 10 minutes. The IRB approved consent form was provided to them for approval. After the subjects were ready with their clothing, the Nexus MK-II was utilized with the electrodes to capture their vital (i.e., heart and respiration signal). For all the subjects, large size clothing was provided at first. The clothing (Single Jersey Knitted Fabric) was fitted using clips, as shown in the following Figure 12 (b).

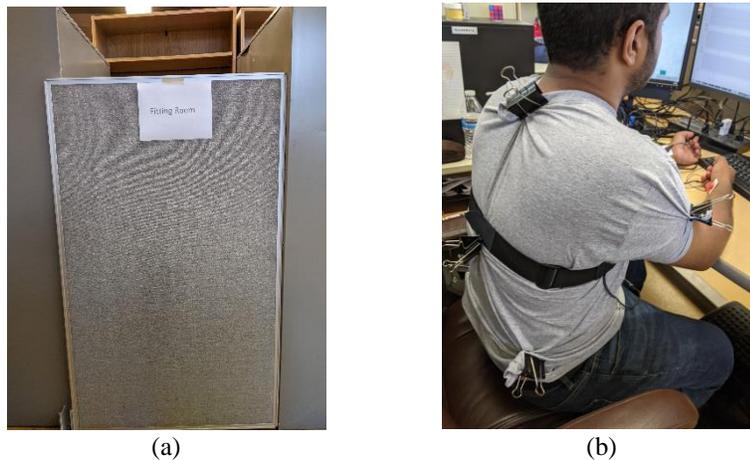

(a) (b)

Figure 12. (a) fitting room, and (b) Heart rate and Respiration rate acquisition using Nexus Mk-II.

*4.2. Comfort parameter selection*

The comfort of clothing has been measured by the physiological parameters previously [4].

Physiological indicators such as heart rate is closely related with user's comfort. For assessment of comfort, heart rate and respiration rate are considered in our study.

*4.2.1. Heart rate vs. fabric type and fit conditions*

Extracted heart rate were used to define a relationship between heart rate and fabric type/ fitting condition.

Table 3. Heart rate of subjects in different fabric and fitting of clothing.

|  | PLF | PTF | CLF | CTF |
|---|---|---|---|---|
| sub1 | 67.66 | 69.18 | 63.73 | 68.26 |
| sub2 | 81.26 | 79.58 | 83.93 | 84.16 |
| sub3 | 72.28 | 73.84 | 72.62 | 75.29 |
| sub4 | 94.23 | 91.97 | 91.97 | 88.27 |
| sub5 | 69.81 | 70.78 | 71.77 | 70.78 |
| sub6 | 71.77 | 73.14 | 74.9 | 69.5 |
| sub7 | 92.53 | 91.42 | 92.53 | 90.35 |
| sub8 | 95.4 | 94.81 | 91.7 | 91.97 |
| sub9 | 91.42 | 91.42 | 91.97 | 92.53 |
| sub10 | 80.41 | 88.27 | 83.02 | 77.96 |
| sub11 | 76.8 | 77.96 | 81.7 | 83.25 |
| Mean | 81.23 | 82.03 | 81.80 | 81.12 |
| Std Dev | 10.51 | 9.68 | 9.92 | 9.21 |
| Variance | 110.57 | 93.77 | 98.50 | 84.92 |

Figure 13 below shows the box whisker plot of the HR distribution of the subject.

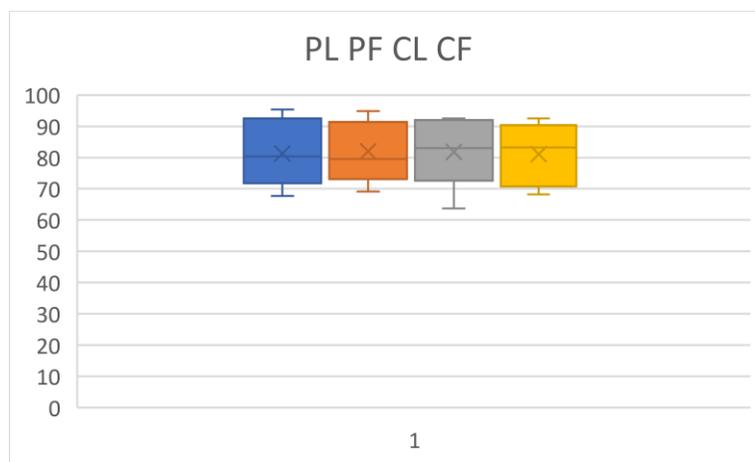

Figure 13. box whisker plot of Heart rate and clothing criteria.

The mean value for each of the clothing criteria doesn't change the heart rate considerably. The student t test for two sample assuming equal variance provides no significant difference for any of the criteria.

Table 4. statistical significance test for heart rate.

|  | **PLF PTF** | **CLF CTF** | **PLF CLF** | **PTF CTF** |
|---|---|---|---|---|
| Pearson Correlation | 0.967475 | 0.948337 | 0.962062 | 0.896792 |
| t stat | -0.98097 | 0.71867 | -0.65748 | 0.701843 |
| P(T<=t) one-tail | 0.174871 | 0.244398 | 0.262859 | 0.249393 |

PLF = Polyester Loose Fit, PTF= Polyester Tight Fit, CLF= Cotton Loose Fit, CTF= Cotton Tight Fit

*4.2.2 Respiration rate vs. fabric type and fit conditions*

Respiration rates from all the 11 subjects were extracted (See Table 5).

Table 5. Respiration data for subjects in different fabric and fit

|  | PLF | PTF | CLF | CTF |
|---|---|---|---|---|
| **sub1** | 15.11 | 15.05 | 15.11 | 13.96 |
| **sub2** | 14.32 | 13.42 | 14.54 | 14.76 |
| **sub3** | 13.33 | 15.80 | 14.38 | 14.06 |
| **sub4** | 14.27 | 12.26 | 14.76 | 12.80 |
| **sub5** | 14.11 | 15.93 | 11.36 | 14.11 |
| **sub6** | 12.26 | 12.22 | 11.60 | 12.84 |
| **sub7** | 10.97 | 11.63 | 11.92 | 13.91 |
| **sub8** | 14.49 | 12.59 | 12.54 | 14.54 |
| **sub9** | 17.61 | 13.81 | 16.13 | 16.99 |
| **sub10** | 14.94 | 14.11 | 13.81 | 13.61 |
| **sub11** | 16.41 | 15.73 | 13.61 | 16.55 |
| **Mean** | 14.35 | 13.86 | 13.61 | 14.37 |
| **Std Dev** | 1.72 | 1.50 | 1.49 | 1.26 |
| **Variance** | 2.97 | 2.26 | 2.23 | 1.61 |

Extracted respiration rate at different fabric type did not show any significant difference. Table 6 shows the data analysis for respiration signals acquired from 11 subjects.

Table 6. statistical significance test for respiration rate.

|  | **PLF PTF** | **CLF CTF** | **PLF CLF** | **PTF CTF** |
|---|---|---|---|---|
| Pearson Correlation | 0.460583 | 0.402891 | 0.666184 | 0.370034898 |
| t stat | 0.659703 | -1.22623 | 0.869098 | -0.812780991 |
| P(T<=t) one-tail | 0.258485 | 0.117178 | 0.19782 | 0.212956761 |

PLF = Polyester Loose Fit, PTF= Polyester Tight Fit, CLF= Cotton Loose Fit, CTF= Cotton Tight Fit

*4.2.3. Assessment of physiological parameter and comfort*

The result is consistent with other previous research (Andreen et. al., 1953; Gavin et. al., 2001; Wingo & McMurry, 2017) in terms of physiological parameters. Although statistical analysis show difference in polyester clothing and cotton clothing, the difference is not significant and needs more data to validate. However, in case of extreme condition (high temperature, humidity, intense exercise) where heart rate and respiration rate could change drastically and physiological comfort parameters are more dominant, our model can predict the fabric type (De Sousa et. al., 2014; Li-ling & Xuan, 2010). Especially, exercise and other extreme parameter (humidity and heat,) different type of fabric (i.e., cotton, wool, viscose) can be suggested for user's comfort (De Sousa et. al., 2014).

*4.3 Emotional response*

Along with quantitative approach of measuring heart rate and respiration rate, this study also implemented a survey-based approach to examine the emotional responses, as another way to measure the comfort. In this study, an online survey was conducted to get the consumer demographic and clothing fit preference data. First, survey questions are to understand the clothing preference and knowledge about the clothing of the participants, in addition to collect their demographic information. In the demographic survey, participants were asked to provide their clothing fit and fabric type preference, the tendency of touching clothes before buying and wearing the garment, as well as preference of fit over weather conditions, comfort, and size. In addition, one open-ended question was asked to see why they prefer the specific fabric type.

Table ##. Demographic information of the participants of the study

| Variables | Category | Responses |
|---|---|---|
| Gender | Male | 55% |
|  | Female | 45% |
| Age | Range | 27-37 |
|  | Mean age | 29.5 |
| Ethnicity | Asian | 90% |
|  | Caucasian | 10% |
| Education | Graduate student | |
| Fit preference (Top) | Fitted | 18% |
| | Semi-fitted | 64% |
| | Loose fitted | 18% |
| Fit preference (Bottom) | Fitted | 9% |
| | Semi-fitted | 82% |
| | Loose fitted | 9% |

A total of 11 participants filled out the survey including the responses about four types of garments used in this study (i.e., cotton-loose fit, cotton-tight fit, polyester-loose fit, polyester-tight fit). For this study, participants were recruited from convenience samples. The detailed demographic information is given in the table## . Regarding the fit preference, the result showed that most of the users preferred semi-fitted garments followed by loosely fitted garments. In the demographic survey data, seven participants also mentioned the priority of choosing a garment is for comfort. They also chose cotton fabrics that are more comfortable to wear probably because it offers of more softness and breathability. On contrary, the style was predominant to other three participants. Interestingly, these three participants also indicated their inclination toward choosing a cotton fabric to wear. The responses from these participants show that they are aware of the comfortability produced from the cotton fabrics. Moreover, one participant mentioned that she would like to wear the garment made of the mix of both cotton and polyester fabrics. According to the participant, the reason for this was that polyester will provide strength in the cloths and cotton will provide soft and breathable features, which shows high-level knowledge about clothing fit and comfortability of the participants. In regard to the question about the tendency of sensing the hand-feel of the fabrics before buying and wearing, 10 out of 11 participants mentioned hand-feel of the fabrics help decide the comfortability of the garment. Based on the participants' responses, weather conditions, size of the garment item, types of fabrics, and style influence the decision of the participants' preference of 'what to wear'.

*Short survey results*

This study also conducted a short survey after wearing cotton and polyester t-shirts by the participants to evaluate their positive and negative emotions. The participants were asked to choose whether they felt soft, comfortable, relaxed, which were considered as positive emotion after wearing a tee-shirt. Items such as Stiff, itchy, annoyed, were considered as the negative feeling about the tee-shirt.

The detailed quantitative data of post-survey after the physiological study showed most of the users are interested in loosely fitted garment regardless of the fabric type (either cotton or polyester) over tightly fitted garment (see Table. ###). The participants preferred loosely fitted garment probably because demography and culture played a vital role in the participant preference that they mostly are from Asia, where loose fit is considered as a traditional part of the fashion (Franks, 2015; Molony, 2007; Tanner, cited from website).

Also, interestingly, between two loosely fitted garments (i.e., cotton vs. polyester), polyester loose fitted garment was more preferred to the participants, which is refuted to the survey responses of the participants saying that they are more likely to wear 100% cotton because of the comfortability, softness, air permeability, and better breathability. Participants also mentioned if they required strong fabric, they would choose a polyester fabric. However, the emotional response data about fabric contradicted the responses of the participants about the type of fabric they found to be comfortable. The probable reasons could be that the study was conducted during summer, where humidity and controlled temperature in the lab environment influenced the participant's decision. The thermal conductivity of polyester fiber is lower than that of cotton fiber (Zhang and Wang, 2018). Therefore, participants might feel comfortable more with a polyester loosely fitted garment over the cotton one because easy cool air was conducted through the wore polyester product because of its lower heat conductivity. Furthermore, the moisture regain of the polyester was lower than the cotton, thus, the polyester might be good to the wearers during the workout or in the sport's relevant activities (Hassan et. al., 2012).

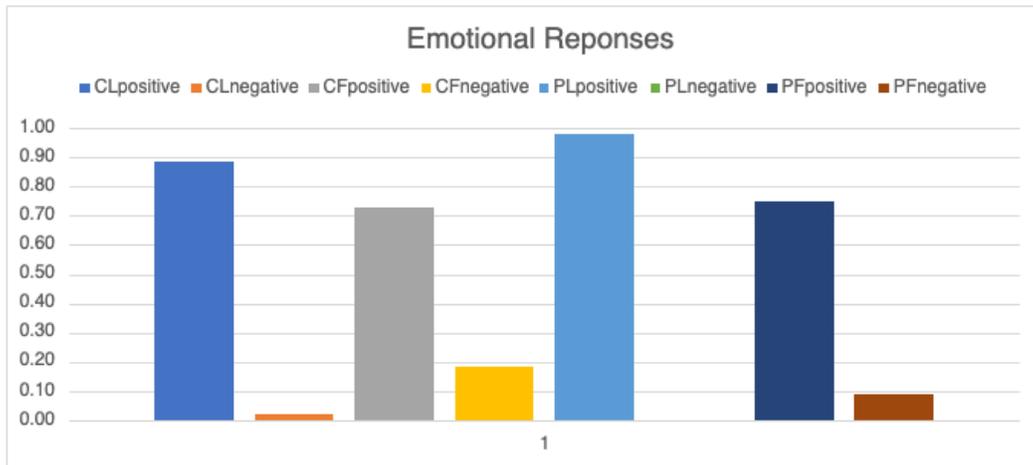

Figure 14. Emotional responses of the participants

## 5. Discussion

This study provided a comprehensive understanding of fabric and fit types of garments that can provide physiological and emotional comfort while wearing those garments. As fit and comfort are important factors affecting consumer's clothing choice, the findings of this study contributed to textile and clothing field by providing a unique way to measure both physiological and emotional parameters for comfort. For the physiological responses, it did not show statistically significant difference in the different fabrics (i.e., cotton and polyester) and fit types (i.e., tight and loose) of the garment. This is probably because the experiment was conducted in a lab setting with a controlled environment (e.g., temperature). However, a small change in respiration and heart rate that might be statistically non-significant statistical could be life-threatening where breathability and moisture absorbency of clothes play a vital role in determining comfort. Although the differences of emotional responses based on the fabric type was not found, the differences were found between the fit type (i.e., loose-fit vs. tight-fit). Since loose-fit garment allows more air circulation and are thermally comfortable, and garments made from natural fibers such as cotton, flax, and hemp products provide more breathability, for daily use, these would be more suitable for the users (Li-Ling and Xuan, 2010). Our finding is in line with the previous research, where researchers showed that non-hygroscopic garments are more prone to producing high heat rate from human body metabolism and sweat from the body (Ha et. al., 1995; Tokura et. al., 1987). From this, for resting and non-laborious works, we recommend natural fiber-based garments and loose-fitted produce to provide more breathable advantages to the human body. However, for laborious works where more sweats might generate, we recommend fabrics that can dry faster, so the human body won't be under risk of developing sweat dermatitis (Soni, et. al., 2019). Furthermore, based on the types of work and physical activities, wearers may choose different types of garments. For example, 100% polyester provides better moisture management for athletes where high heat and sweat are released from the body. Thus, high moisture absorbency and low heat conductivity (such as cotton and other natural fibers) might be harmful to the wearers (Hassan et. al., 2012). For general wearing purposes, several factors wearer needs to consider before selecting any garment including weather, seasons, amount of associated physical activities, and how long time that needs to be worn, among others.

Considering the emotional responses for comfort, the rough surface of the garment might create a more negative feeling of wearing it. Most of the participants in our study mentioned their tendency to touch the fabric to get the hand feel of the garment. Therefore, providing a better tactile experience to the garment could be challenging, but necessary to arouse positive sensory emotions in the consumers. However, excessive dryness in both skin and fabric surfaces might form static electricity as well (Hearle & Morton, 2008). Additional post-processing or treatment may require providing better hand feel comfort and eliminating the possibility of static electricity formation. This also supports the previously conducted research, which showed the necessity of fabric surface modification considering application areas (Sztandera et. al., 2013). This result emphasizes providing information to the consumers regarding the tactile feel of the fabric with currently available information would benefit the wearers.

**Implications**

The finding of this study highlighted the importance of clothing selection based on weather conditions and types of work that wearers intend to do while wearing, so the wearers will get a better heart and respiration rate. Nowadays the availability of different wearables allows for gathering information on health data. Based on the outcomes of this study, different application software is possible to develop that would be linked with the weather data and health data. Subsequently, the app will generate information and recommend the best clothes that fit the users. Furthermore, for the online shoppers, manufacturers may include information about the fabric and fit types along with 'when to wear' information such as weather, type of work, and breathability, among others. The current research finding provided a simplified way of how emotional response data can be useful along with body data to generate generic clothing and health information for the wearers. Further, a multi-disciplinary research approach can disclose more psychological aspects of clothing fit and comfort for different ages, genders, and generations.

*Limitations and future research direction*

One of the limitations of this research is that it was conducted in a lab setting using convenient samples consisting of only 10 people, however, in the future, researchers may recruit more participants from different ethnicity, culture, and background. Moreover, different variables like lab environment, study time and season, and during different works may provide different results. In the future, researchers can conduct a qualitative approach to provide different fabrics feeling and the relation of the fabrics, heart rate, respiration rate, skin irritation, and amount of sweat released from the body wearing those cloths, among others.